\documentclass[prd,aps,nofootinbib,onecolumn,showpacs,12pt]{revtex4-1}
\usepackage{graphicx,epsfig}

\usepackage[table]{xcolor}
\usepackage{epsf}
\usepackage{graphicx}
\usepackage{colortbl}
\definecolor{lightgray}{gray}{0.9}

\begin{document}

\title{  \bf Investigation of the $D^{\ast}_{s}D_{s} \eta^{(\prime)}$ and
$B^{\ast}_{s}B_{s} \eta^{(\prime)}$ vertices via QCD sum rules}
\author{E. Yaz{\i}c{\i} $^{\dag1}$, E. Veli Veliev $^{\dag2}$, K. Azizi $^{*3}$, H. Sundu $^{\dag4}$ \\
 $^{\dag}$Department of Physics , Kocaeli University, 41380 Izmit,
Turkey\\
 $^{*}$Physics Department,  Faculty of Arts and Sciences,
Do\u gu\c s University,
 Ac{\i}badem-Kad{\i}k\"oy, \\ 34722 Istanbul, Turkey\\
$^1$email:enis.yazici@kocaeli.edu.tr\\
$^2$e-mail:elsen@kocaeli.edu.tr\\
$^3$e-mail:kazizi@dogus.edu.tr\\
$^4$email:hayriye.sundu@kocaeli.edu.tr}

\begin{abstract}
The strong coupling constants among mesons  are very important quantities as they can provide 
useful information on  the nature of strong interaction among hadrons  as well as the   QCD
vacuum. In this article, we investigate the strong vertices of  the $D^{\ast}_{s}D_{s} \eta^{(\prime)}$ and
$B^{\ast}_{s}B_{s} \eta^{(\prime)}$ in the framework of the QCD
sum rule approach choosing the $\eta$  or $D_{s} (B_{s})$ meson as an off-shell state. We obtain the results $g_{D_s^*D_s\eta}=(1.46\pm 0.30)GeV^{-1}$,
$g_{D_s^*D_s\eta^{\prime}}=(0.74\pm 0.16)GeV^{-1}$,
$g_{B_s^*B_s\eta}=(5.29\pm 1.06)GeV^{-1}$ and
$g_{B_s^*B_s\eta^{\prime}}=(2.29\pm 0.48)GeV^{-1}$ for the strong coupling constants under consideration, which can be checked in future experiments.
\end{abstract}
\pacs{ 11.55.Hx,  13.75.Lb, 13.25.-k,  13.25.Ft,  13.25.Hw}

\maketitle


\section{Introduction}

In the last few years, both the experimental and theoretical studies on the properties of heavy mesons have received considerable attention. 
With the growing data collected by  many experimental groups, the investigations of the spectroscopy as well as the  electromagnetic, weak and strong decay properties  of the charmed(bottom)--strange mesons have become
 more interesting [1-6]. 
Hence, theoretical determination of various characteristics related to these mesons, such as transition form factors and coupling constants,  become very important for interpretation  of the experimental results.

In the low energy regime of QCD, the large value of the strong coupling constant does  not allow us to use the perturbative theories. Hence, some non-perturbative methods are needed to investigate 
the hadronic properties. The QCD sum rules approach \cite{shifman} is one of the  effective tools in this respect since it is based on QCD Lagrangian and do not  include any model-dependent parameter. 
According to the QCD sum rules method, the strong coupling constants among three mesons are calculated by using three point correlation functions.
In the present work, we apply this technique to investigate the strong coupling constants among $D_s^* [B_s^*]$ and $D_s [B_s]$ 
mesons with light pseudoscalar $\eta$ and $\eta'$ mesons. For some  applications of this method to hadron physics, specially the strong decays see
 \cite{bracco,wang,bracco2,bracco3,bracco4,bracco5,bracco6,bracco7,wang2,bracco8,wang3,maris,gamiz,rosner,lucha,bracco9,bracco10,hollanda,azizi,cui}.

Taking into account only the strong force, the basic SU(3) flavor symmetry for the three light quarks predicts the singlet $\eta_{1}$ and octet $\eta_{8}$ particles:
\begin{equation}
 |\eta_{1}\rangle =\frac{1}{\sqrt{3}} |u\bar{u} + d \bar{d} + s \bar{s}\rangle ,\mbox{ }
 |\eta_{8}\rangle =\frac{1}{\sqrt{6}} |u\bar{u} + d \bar{d} - 2s \bar{s}\rangle
\label{qbase}.
\end{equation}

On the other hand, due to the electromagnetic and weak interactions, a  mixing of these singlet and octet states occurs because of the transformation of  one quark flavor into another.
 The physical $\eta$ and $\eta'$ states are the linear combinations of these SU(3) singlet and octet states:
\begin{equation}
\left(
\begin{array}{c}
\eta \\
\eta'
\end{array}
\right)
 =
\left(
\begin{array}{cc}
\cos\theta &  -\sin\theta \\
\sin\theta  & \cos\theta
\end{array}
\right)
\left(
\begin{array}{c}
\eta_{8} \\
\eta_{1}
\end{array}
\right) , \label{mixq}
\end{equation}
were $\theta$ is the mixing angle of the singlet-octet representation \cite{SVDonskov}. Even though the QCD sum rule is a powerful method for investigation of  the non-perturbative nature of particles,
 the predictions of this approach have considerable uncertainties due to the uncertainties implemented by the quark-hadron duality, determination of the working regions for the Borel mass parameters,
 quark masses, radiative corrections, etc. Hence the relatively small
effect of the  mixing angle allows us to neglect the mixing of the singlet and octet states when QCD sum rules method is used.  In other words, the $\eta$ and $\eta'$ can be taken as pure octet and singlet states, respectively.

The outline of this article is as follows: in section II, we calculate the three point correlation functions for $D^{\ast}_{s}D_{s} \eta^{(\prime)}$ and
$B^{\ast}_{s}B_{s} \eta^{(\prime)}$
 vertices, when  $D_{s}[B_{s}]$ or $\eta [\eta']$ meson is off-shell. Taking into account the quark and  mixed condensate diagrams we obtain QCD sum rules for the strong coupling form factors of each vertex.
 In section III, we present our numerical calculations of the  obtained sum rules and  calculate the values of the strong coupling constants 
for each vertex.

\section{QCD Sum Rules for strong Coupling form factors}

In this section, we obtain  QCD sum rules for strong coupling form factors
associated with the $D^{\ast}_{s}D_{s} \eta[D^{\ast}_{s}D_{s}
\eta^{\prime}]$ and $B^{\ast}_{s}B_{s} \eta[B^{\ast}_{s}B_{s}
\eta^{\prime}]$ vertices by considering the following three-point
correlation functions:
\begin{eqnarray}\label{CorrelationFunc1}
\Pi_{\mu}^{D_s [B_s]}(p^{\prime},q)=i^2 \int d^4x~d^4y~
e^{ip^{\prime}\cdot x}~ e^{iq\cdot y}{\langle}0| {\cal T}\left (
J^{\eta}(x)~ J^{D_s[B_s]}(y)~
J^{D_{s}^{\ast}[B_{s}^{\ast}]\dag}_{\mu}(0)\right)|0{\rangle} ,
\nonumber\\
\Pi_{\mu}^{D_s [B_s]}(p^{\prime},q)=i^2 \int d^4x~d^4y~
e^{ip^{\prime}\cdot x}~ e^{iq\cdot y}{\langle}0| {\cal T}\left (
J^{\eta^{\prime}}(x)~ J^{D_s[B_s]}(y)~
J^{D_{s}^{\ast}[B_{s}^{\ast}]\dag}_{\mu}(0)\right)|0{\rangle}
\end{eqnarray}
for  the case of $D_s[B_s]$ off-shell.  Similarly we consider the correlation  functions for the off-shell  $\eta$ and $\eta^{\prime}$  cases. In Eq. (\ref{CorrelationFunc1}) ${\cal T}$ is  the
time ordering operator and $q=p-p'$ is transferred momentum. The 
 interpolating currents of the participating mesons can be written in terms of the quark
fields as
\begin{eqnarray}\label{mesonintfield}
J^{\eta}(x)&=&\frac{1}{\sqrt{6}}\Big[\overline{u}(x)\gamma_{5}u(x)+\overline{d}(x)\gamma_{5}d(x)-
2\overline{s}(x)\gamma_{5}s(x)\Big] ,
\nonumber \\
J^{\eta^{\prime}}(x)&=&\frac{1}{\sqrt{3}}\Big[\overline{u}(x)\gamma_{5}u(x)+
\overline{d}(x)\gamma_{5}d(x)+
\overline{s}(x)\gamma_{5}s(x)\Big] ,
 \nonumber \\
&&  J^{D_s[B_s]}(x)= \overline{s}(x)\gamma_{5}c[b](x) ,
 \nonumber \\
&&  J^{D_{s}^{\ast}[B_{s}^{\ast}]}_{\mu}(x)=
\overline{s}(x)\gamma_{\mu}c[b](x) .
\end{eqnarray}
The above mentioned correlation functions can be calculated in two different ways. From phenomenological or physical side, they are obtained in terms of hadronic parameters. From theoretical or QCD side,
 they are evaluated in terms of quark's and gluon's degrees of freedom by the help of the operator product expansion (OPE) in deep Euclidean region. After equating the coefficients of individual structures 
from both sides of the same correlation functions, the sum rules for the strong coupling form factors are obtained. Finally we apply
Double Borel transformation with respect to the variables, $p^2$ and $p'^2$  to suppress the contribution of the higher states and continuum. According to the general philosophy of the method, we also 
use the quark-hadron duality assumption.

First, we calculate the physical sides of the
 correlation functions in  Eq. (\ref{CorrelationFunc1}) for the
off-shell $D_s[B_s]$ state. They are obtained by
saturating them  with the complete sets of
appropriate $D_s$, $D_s^{\ast}$ and $\eta^{(\prime)}$ states with
the same quantum numbers as the corresponding mesonic interpolating 
currents. After performing the four-integrals over $x$ and $y$, for both $\eta$ and $\eta^{\prime}$ cases in a compact form, we
get
\begin{eqnarray}\label{CorrelationFuncPhys1}
&&\Pi_{\mu}^{D_s
[B_s]}(p^{\prime},q)\nonumber\\&=&\frac{{\langle}0|J^{\eta^{(\prime)}}|\eta^{(\prime)}
(p^{\prime}){\rangle} {\langle}0|J^{D_s[B_s]}|D_s[B_s]
(q){\rangle} {\langle}\eta^{(\prime)}(p^{\prime}) D_s[B_s]
(q)|D_{s}^{\ast}[B_{s}^{\ast}](p,\epsilon){\rangle}
{\langle}D_s^{\ast}[B_{s}^{\ast}]
(p,\epsilon)|J^{D_s^{\ast}[B_{s}^{\ast}]}_{\mu}|0{\rangle}}
{(q^2-m_{D_s[B_s]}^2)
(p^2-m_{D_s^{\ast}[B_{s}^{\ast}]}^2)({p^{\prime}}^{2}-m_{\eta^{(\prime)}}^2)}
\nonumber \\ &+&...  ,
\end{eqnarray}
where .... stands for the contributions of the higher states and
continuum. To proceed we need to define the following  matrix elements  in terms of hadronic parameters:
\begin{eqnarray}\label{transitionamp}
{\langle}
0|J^{\eta^{(\prime)}}|\eta^{(\prime)}(p^{\prime}){\rangle}&=&\frac{m_{\eta^{(\prime)}}^2f_{\eta^{(\prime)}}}{2m_s} ,
\nonumber\\
{\langle}0|J^{D_s[B_s]}|D_s[B_s](q){\rangle}&=&\frac{m_{D_s[B_s]}^2~
f_{D_s[B_s]}}{m_{c(b)}+m_s} ,
\nonumber\\
{\langle}D_{s}^{\ast}[B_{s}^{\ast}](p,\epsilon)
 |J^{D_{s}^{\ast}[B_{s}^{\ast}]}_{\mu}|0{\rangle}&=&m_{D_{s}^{\ast}[B_{s}^{\ast}]}
 f_{D_{s}^{\ast}[B_{s}^{\ast}]}{\epsilon^{\ast}}_{\mu} ,
 \nonumber\\
{\langle}\eta^{(\prime)}(p^{\prime})D_s[B_s](q)|D^{\ast}_{s}[B^{\ast}_{s}](p,\epsilon){\rangle}
&=&g_{D^{\ast}_{s}D_s \eta^{(\prime)}[B^{\ast}_{s}B_s
\eta^{(\prime)}]}^{D_s[B_s]}(p^{\prime}-q)\cdot \epsilon ,
\end{eqnarray}
where $g_{D^{\ast}_{s}D_s \eta^{(\prime)}[B^{\ast}_{s}B_s
\eta^{(\prime)}]}^{D_s[B_s]}$ is the strong coupling form factor; and $f_{D_s^{\ast}[B_s^{\ast}]}$, $f_{D_s[B_s]}$ and
$f_{\eta^{(\prime)}}$ are leptonic decay constants of the
$D_s^{\ast}[B_s^{\ast}]$, $D_s[B_s]$ and $\eta^{(\prime)}$ mesons, respectively. Using
Eq. (\ref{transitionamp}) in Eq. (\ref{CorrelationFuncPhys1}) and
summing over polarization vectors, we obtain 
the physical side  as
\begin{eqnarray}\label{CorrelationFuncPhys2}
\nonumber\\
\Pi_{\mu}^{D_s[B_s]}(p^{\prime},q)&=&g^{D_s[B_s]}_{D^{\ast}_{s}D_s
\eta^{(\prime)}[B^{\ast}_{s}B_s
\eta^{(\prime)}]}\frac{f_{D_s^{\ast}[B_s^{\ast}]}m_{D_s^{\ast}[B_s^{\ast}]}
f_{\eta^{(\prime)}}m_{\eta^{(\prime)}}^2
f_{D_s[B_s]}m_{D_s[B_s]}^2}
{(q^2-m_{D_s[B_s]}^2)(p^{\prime^2}-m_{\eta^{(\prime)}}^2)(p^2-m_{D_s^{\ast}(B_s^{\ast})}^2)2m_{s}(m_{c(b)}+m_s)}
\nonumber\\
&&\times \Big[\Big(1+\frac{m_{\eta^{(\prime)}}^2-q^2}{m_{D_s^{\ast}[B_s^{\ast}]}^2}\Big)p_{\mu}-
2p^{\prime}_{\mu}\Big] + ....,
\nonumber\\
\end{eqnarray}
where we will choose the structure  $p_{\mu}$ to calculate the corresponding strong coupling  form factor.  From a similar manner, one can obtain the final expression of the physical side
of the correlation function for an  $\eta^{(\prime)}$ off-shell.

\begin{figure}[h!]
\begin{center}
\includegraphics[width=17cm]{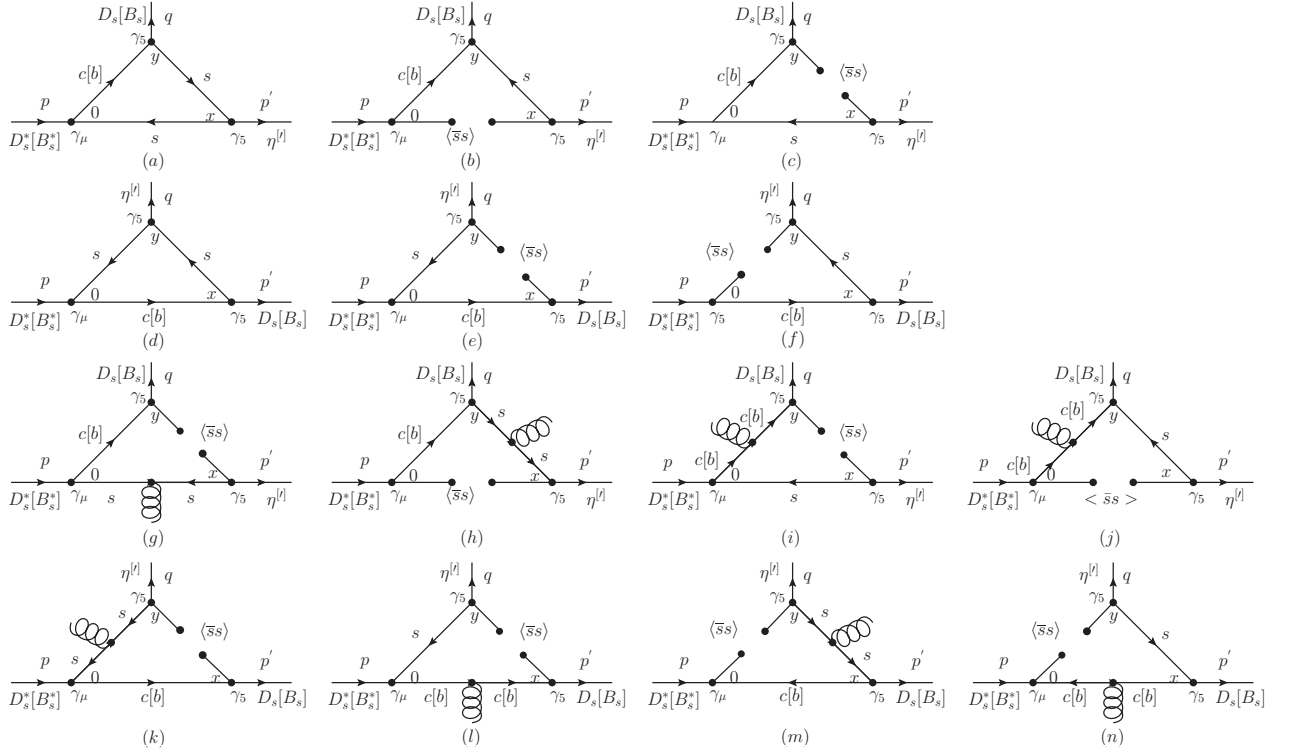}
\end{center}
\caption{Diagrams considered in the calculations.}
\label{Figure}
\end{figure}

From the QCD or theoretical side,   the aforesaid correlation
functions are calculated in deep Euclidean space, where $p^2\rightarrow-\infty$
and ${p^{\prime}}^2\rightarrow-\infty$ by the help of OPE. To obtain the QCD representation, as an example for the $D_s[B_s]$ off-shel case, we separate the correlation function into 
perturbative and non-perturbative parts and keep only the structure which we use to extract the sum rules
\begin{eqnarray}\label{CorrelationFuncQCD}
\Pi^{D_s[B_s]}_{\mu}(p^{\prime},q)&=& \left(\Pi_{per}+\Pi_{nonper}\right)p_{\mu}+...,
\end{eqnarray}
where  the  perturbative part can be expressed in terms of a double dispersion integral of the form
\begin{eqnarray}\label{CorrelationFuncQCDPert}
\Pi_{per}&=&-\frac{1}{4 \pi^{2}} \int ds \int ds^{\prime}
\frac{\rho(s,s^{\prime}, q^2)}{(s-p^2)
(s^{\prime}-{p^{\prime}}^2)}+\mbox{subtraction terms},
\end{eqnarray}
with $\rho(s,s^{\prime}, q^2) $ being the corresponding spectral density. Our main task in the following is  to calculate  this spectral density. For this aim, we  consider the 
bare loop diagrams (a) and (d) in Fig. \ref{Figure} for $D_s[B_s]$  as off-shell state. We
calculate these diagrams via Cutkosky rules, as a result of which we get
\begin{eqnarray}\label{SpecDenstD}
\rho^{D_s[B_s]}(s,s^{\prime},q^2)&=&-\frac{N_c}{\lambda^{3/2}(s,s^{\prime},q^2)}
\left\{m_{c[b]}^2s^{\prime}(q^2-s+s^{\prime})-m_s^2s^{\prime}(q^2-s+s^{\prime})+q^2s^{\prime}
(-q^2+s+s^{\prime})\right\}.\nonumber\\
\end{eqnarray}
Similarly, for the case of $\eta^{(\prime)}$ off-shel one gets
\begin{eqnarray}
\rho^{\eta^{(\prime)}}(s,s^{\prime},q^2)&=&\frac{N_c}{\lambda^{3/2}(s,s^{\prime},q^2)}
\left\{m_s^2\Big(q^4-2q^2s+(s-s^{\prime})^2\Big)
-m_{c[b]}m_s\Big(q^4+(s-s^{\prime})^2-2q^2(s+s^{\prime})\Big)\right.
\nonumber \\
 & -&\left. 2m_{c[b]}^2q^2s^{\prime}(s+s^{\prime}-q^2) \right\},
\end{eqnarray}
where $\lambda(a,b,c)=a^2+b^2+c^2-2ac-2bc-2ab$ and
$N_c=3$ is the color number.

To calculate the non-perturbative contributions in QCD side, we
consider  all condensate diagrams in Fig. \ref{Figure}.  As a result, we get
\begin{eqnarray}\label{CorrelationFuncNonpert}
\Pi_{nonper}^{D_s[B_s]}&=&
{\langle}\overline{s}s{\rangle}m_0^2
\Big[\frac{m_{c[b]}+2m_s}{12r^3}+\frac{m_s^3}{2rr^{\prime^3}}+
\frac{m_{c[b]}^2m_s+m_s^3-m_s
q^2}{4r^2r^{\prime^2}}+\frac{3m_s}{4rr^{\prime^2}}+\frac{m_{c[b]}+2m_s}{12r^2r^{\prime}}
\nonumber \\
 & +& \frac{5m_{c[b]}^2m_s-2m_{c[b]}^3+m_{c[b]}m_s^2+2m_s^3+2m_{c[b]}q^2
 -2m_sq^2}{12r^3r^{\prime}}\Big]+{\langle}\overline{s}s{\rangle}\Big[\frac{m_s}{2r^2}-\frac{m_s^5}{rr^{\prime^3}}
 \nonumber \\&-&\frac{m_{c[b]}^2m_s^3}{r^3r^{\prime}}-\frac{m_s}{rr^{\prime}}
+
 \frac{m_{c[b]}^2m_s-2m_{c[b]}m_s^2+m_s^3-m_sq^2}{2r^2r^{\prime}}
 +\frac{m_s^3q^2-m_s^3m_{c[b]}^2-m_s^5}{2r^2r^{\prime^2}}\Big],\nonumber \\
\end{eqnarray}
for the case of $D_s[B_s]$  off-shell and
\begin{eqnarray}\label{CorrelationFuncNonpert}
\Pi_{nonper}^{\eta^{(\prime)}}&=&0,
\end{eqnarray}
for the case of $\eta^{(\prime)}$  off-shell, where
$r=p^2-m_{c[b]}^2$ and $r^{\prime}=p^{\prime^2}-m_s^2$.

As we previously mentioned, the sum rules for strong coupling form factors are obtained by equating the coefficients of the selected structure from 
phenomenological and QCD sides of the correlation functions and applying
double Borel transformation as well as continuum subtraction. After these procedures, we obtain 
\begin{eqnarray}\label{CoupCons-gDsDKs-Doffshel}
&&g^{D_s[B_s]}_{D^{\ast}_{s}D\eta^{(\prime)}[B^{\ast}_{s}B\eta^{(\prime)}]}(q^2)\nonumber\\&=&
\frac{2m_s(m_{c[b]}+m_s)
(q^2-m_{D_s[B_s]}^2)}{f_{D_s^{\ast}[B_s^{\ast}]} f_{D_s[B_s]}
f_{\eta^{(\prime)}}m_{D_s^{\ast}[B_s^{\ast}]}m^2_{\eta^{(\prime)}}m_{D_s[B_s]}^2}\Big(1+\frac{m_{\eta^{(\prime)}}^2-q^2}
{m_{D_s^{\ast}[B_s^{\ast}]}^2}\Big)^{-1}
exp\Big[\frac{m_{D_s^{\ast}[B_s^{\ast}]}^2}{M^2}\Big]exp\Big[\frac{m_{\eta^{(\prime)}}^2}{{M^{\prime}}^2}\Big]
\nonumber \\
&& \times \left[\frac{1}{4~\pi^2}\int^{s_0}_{(m_c+m_s)^2}
ds\int^{s^{\prime}_0}_{4m_s^2} ds^{\prime}
\rho^{D_s[B_s]}(s,s^{\prime},q^2) \theta
[1-{(f^{D_s[B_s]}(s,s^{\prime}))}^2]e^{\frac{-s}{M^2}}e^{\frac{-s^{\prime}}
{{M^{\prime}}^2}}+\widehat{\mathbf{B}}\Pi_{nonper}^{D_s[B_s]}\right],
\nonumber\\
\end{eqnarray}
where $M^2$ and ${M^{\prime}}^2$ are Borel mass parameters and $s_0$ and $s^{\prime}_0$ are continuum thresholds. The function $\widehat{\mathbf{B}}\Pi_{nonper}^{D_s[B_s]}$ is given by
\begin{eqnarray}\label{BorelDs-ofshell}
&&\widehat{\mathbf{B}}\Pi_{nonper}^{D_s[B_s]}\nonumber\\&=&e^{\frac{-m_{c[b]}^2}{M^2}}e^{\frac{-m_s^2}
{{M^{\prime}}^2}}{\langle}\overline{s}s{\rangle}\Big\{m_0^2\Big(
\frac{m_s^3}{4M^{\prime^4}}-\frac{3m_s}{4M^{\prime^2}}-\frac{m_{c[b]}+2m_s}{12M^2}+\frac{m_{c[b]}q^2-m_sq^2}{12M^4}+\frac{m_{c[b]}^2m_s+m_s^3-m_{s}q^2}{4M^2M^{\prime^2}}
\nonumber \\
&-&
\frac{2m_{c[b]}^3-5m_{c[b]}^2m_s-m_{c[b]}m_s^2-2m_s^3}{24M^4}\Big)
-m_s-\frac{m_s^5}{2M^{\prime^4}}
\nonumber \\
&-&\frac{m_{c[b]}^2m_s
+2m_{c[b]}m_s^2-m_s^3+m_sq^2}{2M^2}-
\frac{m_{c[b]}^2m_s^3}{2M^4}-\frac{m_{c[b]}^2m_s^3+m_s^5-m_s^3q^2}{2M^2M^{\prime^2}}\Big\},
\end{eqnarray}
for the off-shell $D_s[B_s]$ state
and
\begin{eqnarray}\label{CoupCons-gDsDKs-Ksoffshel}
&&g^{\eta^{(\prime)}}_{D^{\ast}_{s}D\eta^{(\prime)}[B^{\ast}_{s}B\eta^{(\prime)}]}(q^2)\nonumber\\&=&
\frac{2m_s(m_{c[b]}+m_s)
(q^2-m_{\eta^{(\prime)}}^2) m_{D_s^{\ast}[B_s^{\ast}]}^2}{f_{D_s^{\ast}[B_s^{\ast}]}f_{D_s[B_s]}f_{\eta^{(\prime)}}
m_{D_s^{\ast}[B_s^{\ast}]}m^2_{\eta^{(\prime)}}
m_{D_s[B_s]}^2 (m_{D_s^{\ast}[B_s^{\ast}]}^2+m_{D_s[B_s]}^2-q^2)}exp\Big[\frac{m_{D_s^{\ast}[B_s^{\ast}]}^2}{M^2}\Big]exp\Big[\frac{m_{D_s[B_s]}^2}{{M^{\prime}}^2}\Big]
\nonumber \\
&& \times \Big[\frac{1}{4~\pi^2}\int^{s_0}_{(m_c+m_s)^2}
ds\int^{s^{\prime}_0}_{(m_c+m_s)^2} ds^{\prime}
\rho^{\eta^{(\prime)}}(s,s^{\prime},q^2)
\theta[1-(f^{\eta^{(\prime)}}(s,s^{\prime}))^2]e^{\frac{-s}{M^2}}e^{\frac{-s^{\prime}}
{M^{\prime^2}}}\Big],
\end{eqnarray}
for the   $\eta^{(\prime)}$ off-shell  case.

The functions $f^{D_s[B_s]}(s,s^{\prime})$ and $f^{\eta^{(\prime)}}(s,s^{\prime})$ in the above equations are defined as
\begin{eqnarray}\label{fsspD0offshell}
 f^{D_s[B_s]}(s,s^{\prime})&=&\frac{2~s~s^{\prime}+(m_{c[b]}^2-m_s^2-s)
(-q^2+s+s^{\prime})}{\lambda^{1/2}(m_{c[b]}^2,m_s^2,s)
\lambda^{1/2}(s,s^{\prime},q^2)},\nonumber\\
f^{\eta^{(\prime)}}(s,s^{\prime})&=&\frac{2~s~(-m_{c[b]}^2+m_s^2-s^{\prime})+(m_{c[b]}^2-m_s^2+s)
(-q^2+s+s^{\prime})}{\lambda^{1/2}(m_{c[b]}^2,m_s^2,s)
\lambda^{1/2}(s,s^{\prime},q^2)}.
\end{eqnarray}

\section{Numerical results}

In this section we numerically analyze the sum
rules obtained in the previous section to obtain the behavior of the strong coupling form factors in terms of $q^2$. For this purpose we use some input parameters listed in Table I.
\begin{table}[ht]\label{table1}
\centering \rowcolors{1}{lightgray}{white}
\begin{tabular}{cc}
\hline \hline
   Parameters  &  Values
           \\
\hline \hline
$m_{c}$              & $ (1.275\pm 0.025)~ GeV~$\cite{Beringer}\\
$m_{b}$              & $(4.65\pm0.03)~ GeV~$\cite{Beringer}\\
$ m_{s} $              &   $ (95\pm5)~MeV~$\cite{Beringer} \\
$ m_{B_s^*} $            &   $ (5415.4^{+2.4}_{-2.1})~MeV~$\cite{Beringer} \\
$ m_{B_s}$           &   $ (5366.77\pm0.24)~MeV~$\cite{Beringer} \\
$ m_{D_s^*}$    &   $ (2112.3\pm0.5)~MeV~$\cite{Beringer}   \\
$ m_{D_s} $      &   $ (1968.49\pm0.32)~MeV~$\cite{Beringer}   \\
$ m_{\eta} $      &   $ (547\pm 0.024)~ MeV~ $\cite{Beringer}   \\
$ m_{\eta^{\prime}} $      &   $(958\pm 0.06)~ MeV~$\cite{Beringer}   \\
$ f_{B_s^*} $      &   $(0.229)~ GeV~ $\cite{becirevic}   \\
$ f_{B_s} $      &   $(0.196)~ GeV~ $\cite{ivanov}   \\
$ f_{D_s^*}$            &   $(0.272)~ GeV~$\cite{becirevic} \\
$  f_{D_s}$ &   $ (0.286)~ GeV~ $\cite{abbiendi}   \\
$ f_{\eta} $      &   $(0.174)~ GeV~ $\cite{Yeni}   \\
$ f_{\eta^{\prime}} $      &   $(0.170)~ GeV~ $\cite{Yeni}   \\
$ \langle0|\overline{s}s(1GeV)|0\rangle$        &   $ -0.8(0.24\pm0.01)^3~GeV^{3}~$\cite{Ioffe}   \\
$ m_0^2(1GeV) $       &   $(0.8\pm0.2)$ $GeV^{2}$   \\
 \hline \hline
\end{tabular}
\caption{Input parameters used in our calculations.}
\end{table}

The sum rules for the form factors contain also four auxiliary parameters: Borel
mass parameters $M^2$ and ${M^{\prime}}^2$as well as continuum thresholds  $s_0$ and
$s_0^{\prime}$. In the following, we proceed to find  working regions for these auxiliary parameters  at which the dependences of coupling
form factors on these  parameters are weak. The working
regions for the Borel  parameters $M^2$ and ${M^{\prime}}^2$
are calculated demanding that both the contributions of the higher
states and continuum are adequately suppressed and the
contributions of the higher dimensional operators are small. These conditions lead to the regions $5~GeV^2\leq M^2\leq9~GeV^2$ and
 $2~GeV^2\leq M'^2\leq7~GeV^2$ for $D_s$ as off-shell meson, as well as $5~GeV^2\leq M^2\leq9~GeV^2$ and
 $5~GeV^2\leq M'^2\leq9~GeV^2$ for $\eta^{(\prime)}$ off-shell associated with the
$D_s^{\ast}D_s \eta^{(\prime)}$ vertex. We also find  the regions
$20~GeV^2\leq M^2\leq30~GeV^2$ and
 $3~GeV^2\leq M'^2\leq6~GeV^2$ for $B_s$ off-shell, as well as $10~GeV^2\leq M^2\leq20~GeV^2$ and
 $10~GeV^2\leq M'^2\leq20~GeV^2$ for $\eta$ off-shell in accordance with  the $B_s^{\ast}B_s
 \eta$
 vertex. For the $B_s^{\ast}B_s\eta^{\prime}$ vertex the regions
$10~GeV^2\leq M^2\leq20~GeV^2$ and
 $3~GeV^2\leq M'^2\leq6~GeV^2$ for $B_s$ off-shell, as well as  $10~GeV^2\leq M^2\leq20~GeV^2$ and
 $10~GeV^2\leq M'^2\leq20~GeV^2$ for the case of  $\eta{\prime}$ off-shell  are obtained.

The continuum thresholds $s_0$ and
$s_0^{\prime}$ are not totally arbitrary  but they are related  to the
energy of the first excited states in initial and final channels with the same quantum numbers.
Our numerical analysis leads to the following working regions for the
continuum thresholds in $s$ and $s'$ channels for different off-shel cases and vertecies:
$(m_{D_s^{\ast}[B_s^{\ast}]}+0.3)^2\leq s_0
\leq(m_{D_s^{\ast}[B_s^{\ast}]}+0.7])^2$ for all off-shell cases
in $s$ channel,  $(m_{D_s[B_s]}+0.3)^2\leq s_0^{\prime}\leq
(m_{D_s[B_s]}+0.7)^2$ for $D_s[B_s]$ off-shell and
$(m_{\eta^{(\prime)}}+0.3)^2\leq s_0^{\prime}\leq
(m_{\eta^{(\prime)}}+0.5)^2$ for $\eta^{(\prime)}$ 
off-shell in $s'$ channel. 

\begin{figure}[h!]
\includegraphics[totalheight=7cm,width=8cm]{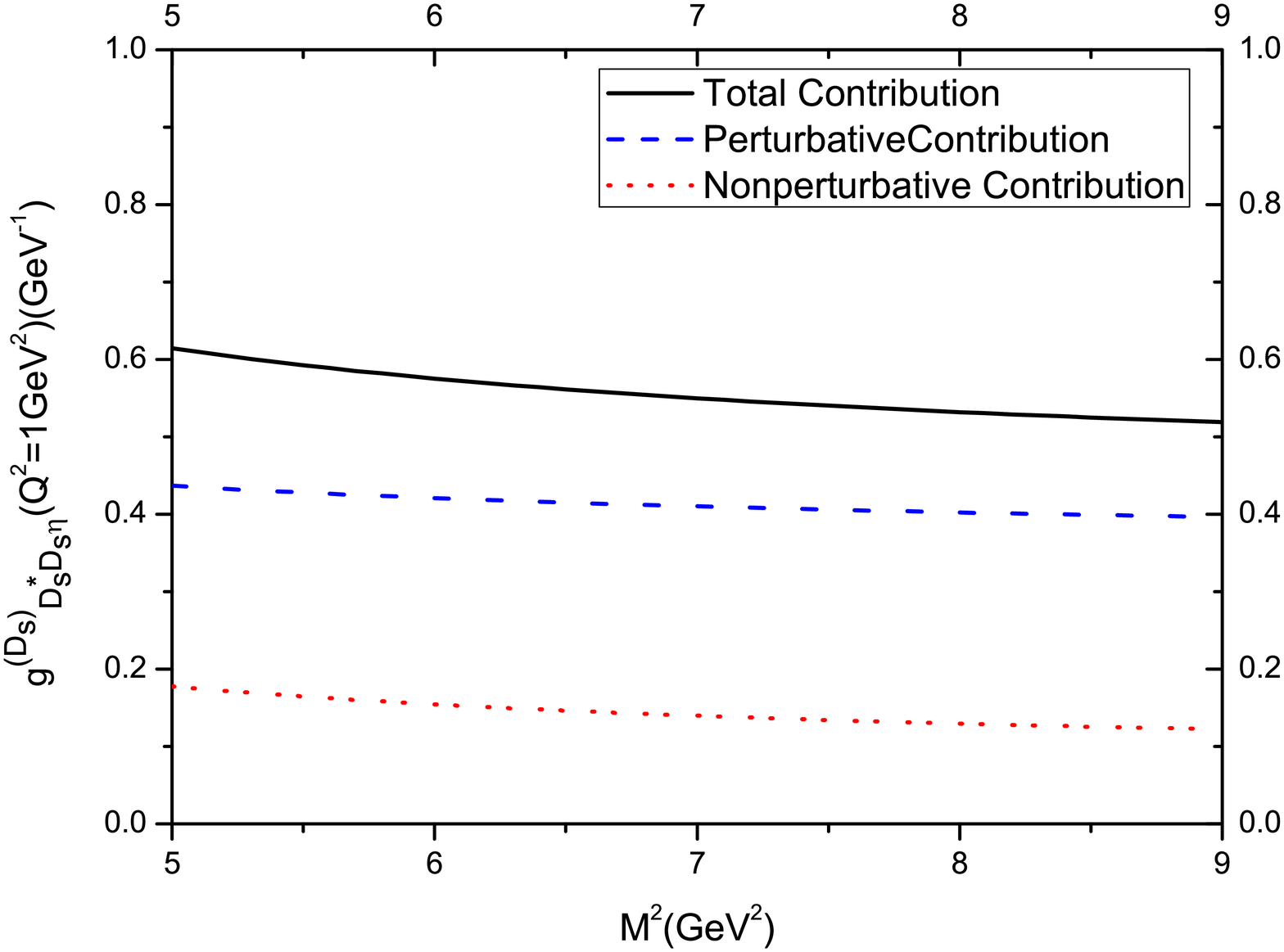}
\includegraphics[totalheight=7cm,width=8cm]{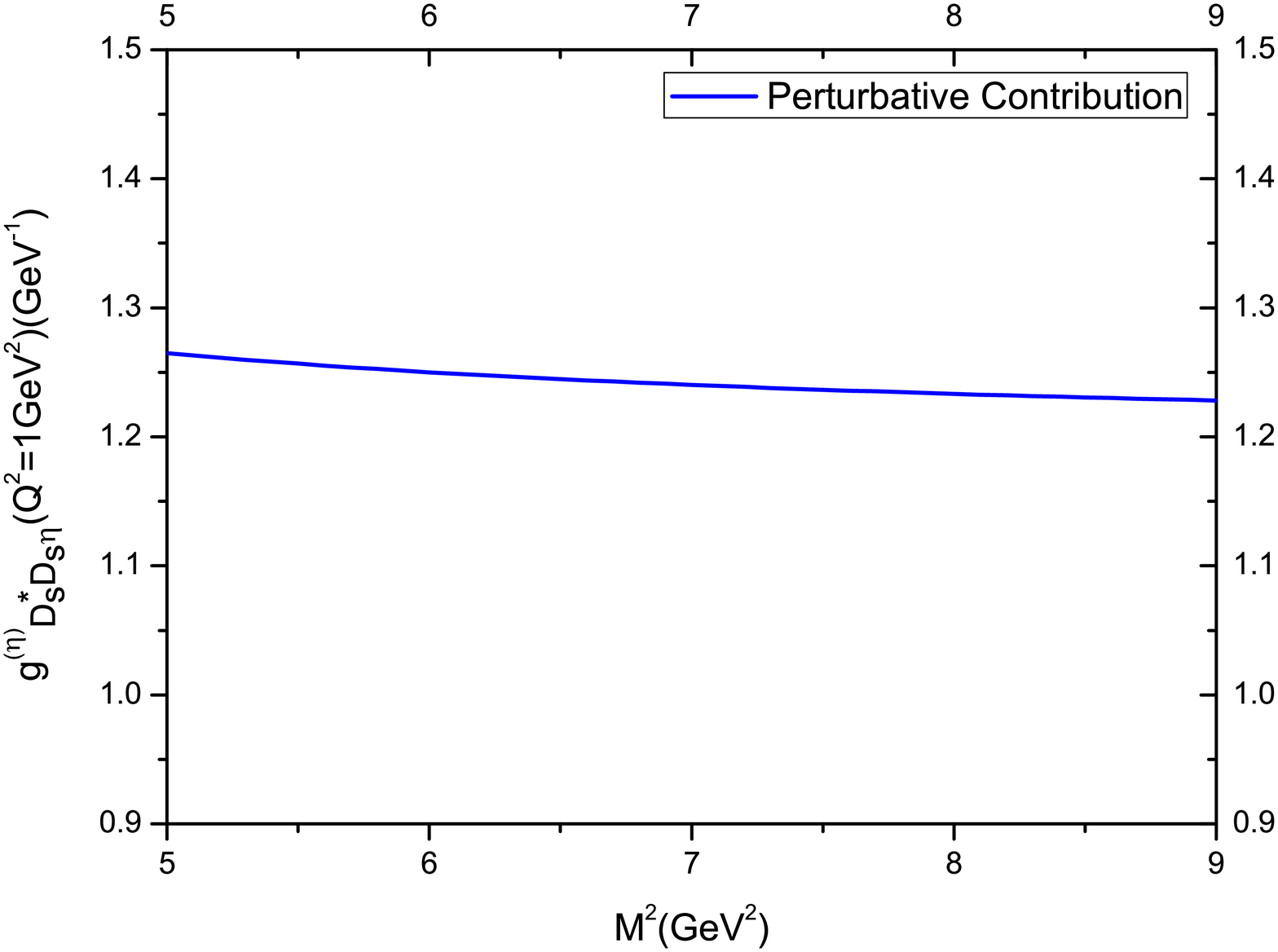}
\caption{\textbf{Left:} $g^{(D_s)}_{D_s^*D_s\eta}(Q^2=1~GeV^2)$ as
a function of the Borel mass parameter $M^2$. \textbf{Right:}
$g^{(\eta)}_{D_s^*D_s\eta}(Q^2=1~GeV^2)$ as a function of the
Borel mass $M^2$. } \label{gDssDseta}
\end{figure}
\begin{figure}[h!]
\includegraphics[totalheight=7cm,width=8cm]{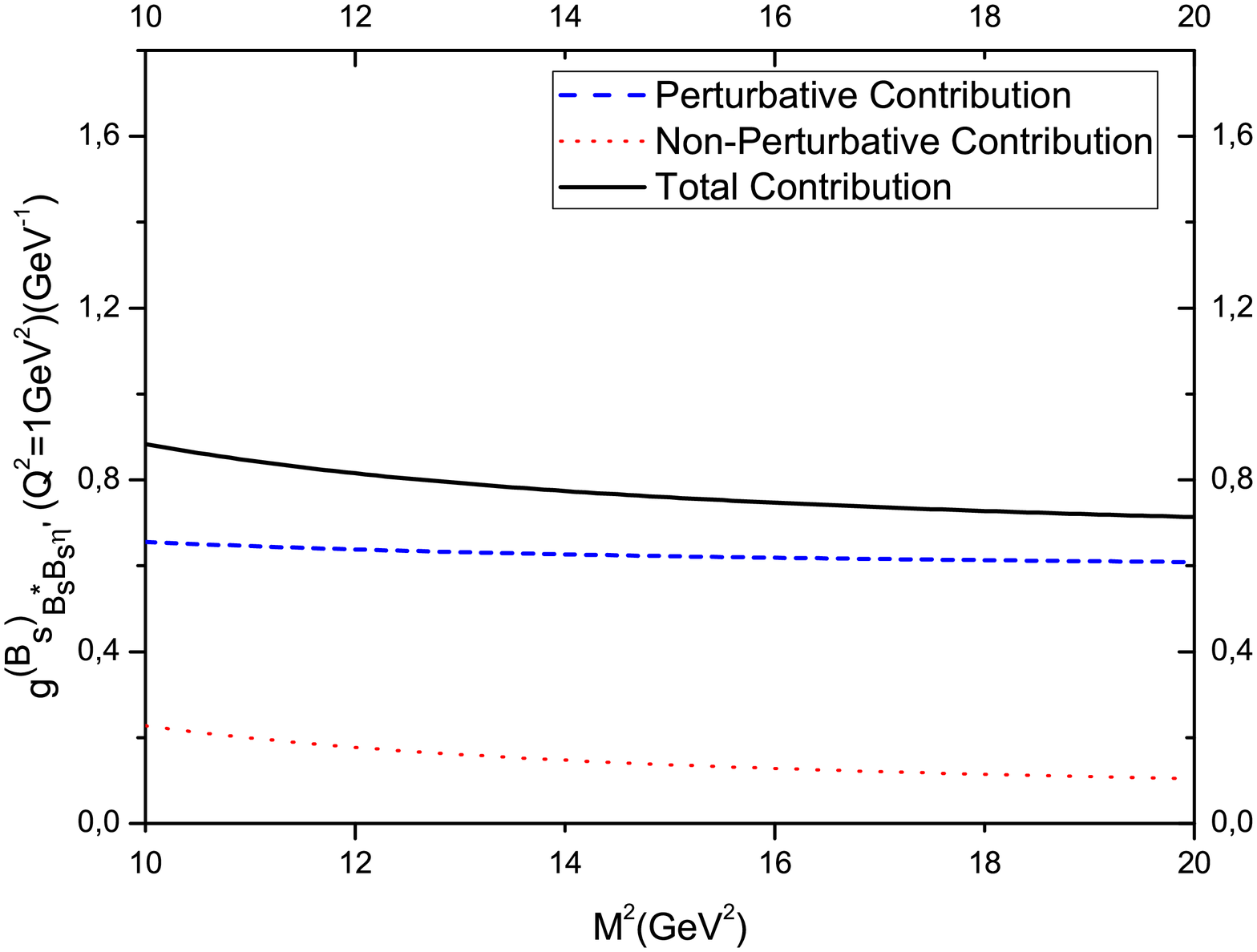}
\includegraphics[totalheight=7cm,width=8cm]{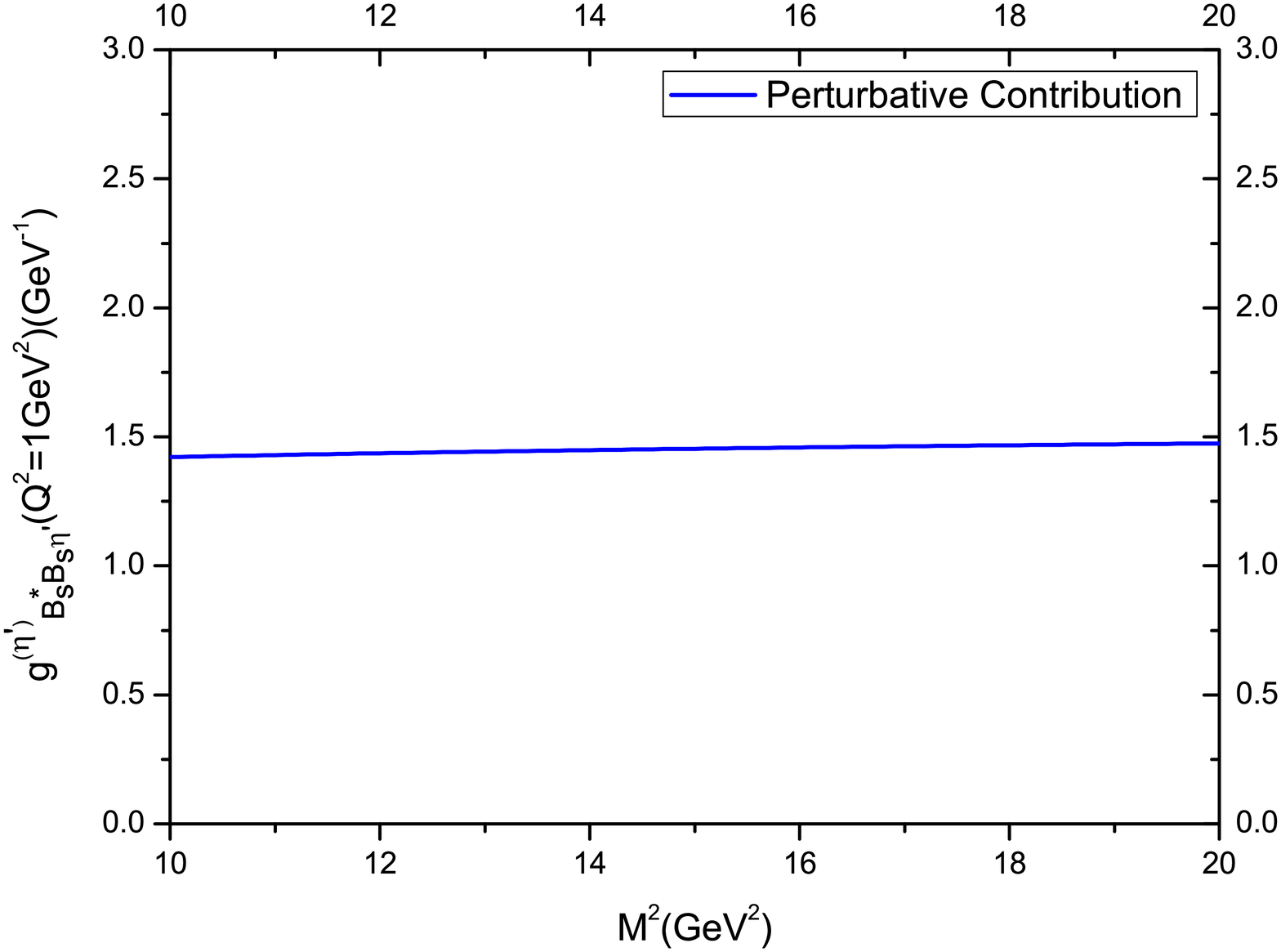}
\caption{\textbf{Left:}
$g^{(B_s)}_{B_s^*B_s\eta^{\prime}}(Q^2=1~GeV^2)$ as a function of
the Borel mass $M^2$. \textbf{Right:}
$g^{(\eta^{\prime})}_{B_s^*B_s\eta^{\prime}}(Q^2=1~GeV^2)$ as a
function of the Borel mass parameter $M^2$. } \label{gBssBseta}
\end{figure}

 Having determined the working regions for auxiliary parameters,  we present the dependences of some strong form factors under consideration at $Q^2=-q^2=1~GeV^2$ for instance on Borel
parameter $M^2$ for different off-shell cases  in Figs. \ref{gDssDseta}
and \ref{gBssBseta}. From these figures, we see that the strong form factors depict good
stabilities  with respect to the variations of the $M^2$ in its working regions. 
By using the working regions for all  auxiliary parameters and other
inputs, we obtain that the strong 
form factors are well fitted to  the following function (see figure 4):
\begin{eqnarray}\label{FitFunc}
g(Q^2)=\alpha+\gamma\exp[-\beta~Q^2] ,
\end{eqnarray}
where the values of the parameters $\alpha$, $\beta$
and $\gamma$ for different cases are
given in Table \ref{fitparametrization0}.

\begin{table}[h]
\renewcommand{\arraystretch}{1.5}
\addtolength{\arraycolsep}{3pt}
$$
\begin{array}{|c|c|c|c|c|}
\hline \hline
 \mbox{ } &
\alpha (GeV^{-1})& \gamma (GeV^{-1}) &  \beta (GeV^{-2})
 \\
\hline\hline
  \mbox{$g^{(D_s)}_{D_s^*D_s\eta}(Q^2)$}
      & 0.3682 & 0.2383 & 0.2753
      \\
      \hline
  \mbox{$g^{(\eta)}_{D_s^*D_s\eta}(Q^2)$}
      & -0.7125 & 2.4431 & 0.1791
      \\
      \hline
  \mbox{$g^{(D_s)}_{D_s^*D_s\eta^{\prime}}(Q^2)$}
      & 0.2216 & 0.2768 & 0.1888
      \\
       \hline
  \mbox{$g^{(\eta^{\prime})}_{D_s^*D_s\eta^{\prime}}(Q^2)$}
      & -0.0869 & 0.7132 & 0.3000
      \\
      \hline
  \mbox{$g^{(B_s)}_{B_s^*B_s\eta}(Q^2)$}
      & 1.2167 & 0.4576 & 0.0688
      \\
       \hline
  \mbox{$g^{(\eta)}_{B_s^*B_s\eta}(Q^2)$}
      & -7.2443 & 12.9086 & 0.0965
      \\
       \hline
  \mbox{$g^{(B_s)}_{B_s^*B_s\eta^{\prime}}(Q^2)$}
      & 0.4990 & 0.2762 & 0.0601
      \\
      \hline
  \mbox{$g^{(\eta^{\prime})}_{B_s^*B_s\eta^{\prime}}(Q^2)$}
      & -0.6228 & 2.6885 & 0.2407
      \\
    \hline \hline
\end{array}
$$
\caption{Parameters appearing in the fit function of the coupling
constants. } \label{fitparametrization0}
\renewcommand{\arraystretch}{1}
\addtolength{\arraycolsep}{-1.0pt}
\end{table}

The coupling constants are defined as the values of the strong  form
factors at $Q^2=-m_{off-shell}^2$. The numerical results of the coupling constants for different vertecies are
given in Table \ref{CouplingConstNumericValue}. The final result
for each coupling constant is obtained by taking the average of
the coupling constants obtained from two different off-shell
cases, which also are presented in Table \ref{CouplingConstNumericValue}. The errors in
the numerical values of the strong coupling constants  are due to the uncertainties in
determination of the working regions for the auxiliary parameters
as well as the errors in other input parameters. 

In summary, we calculated the strong coupling form factors of  the $D^{\ast}_{s}D_{s} \eta^{(\prime)}$ and
$B^{\ast}_{s}B_{s} \eta^{(\prime)}$ vertices for different off-shell cases in the frame work of the QCD sum rules. By obtaining the behavior of the strong form factors in terms of $Q^2$, we
also calculated the strong coupling constants corresponding to the considered vertices. Our predictions can be checked in future experiments. 

\begin{figure}[h!]
\includegraphics[totalheight=7cm,width=8cm]{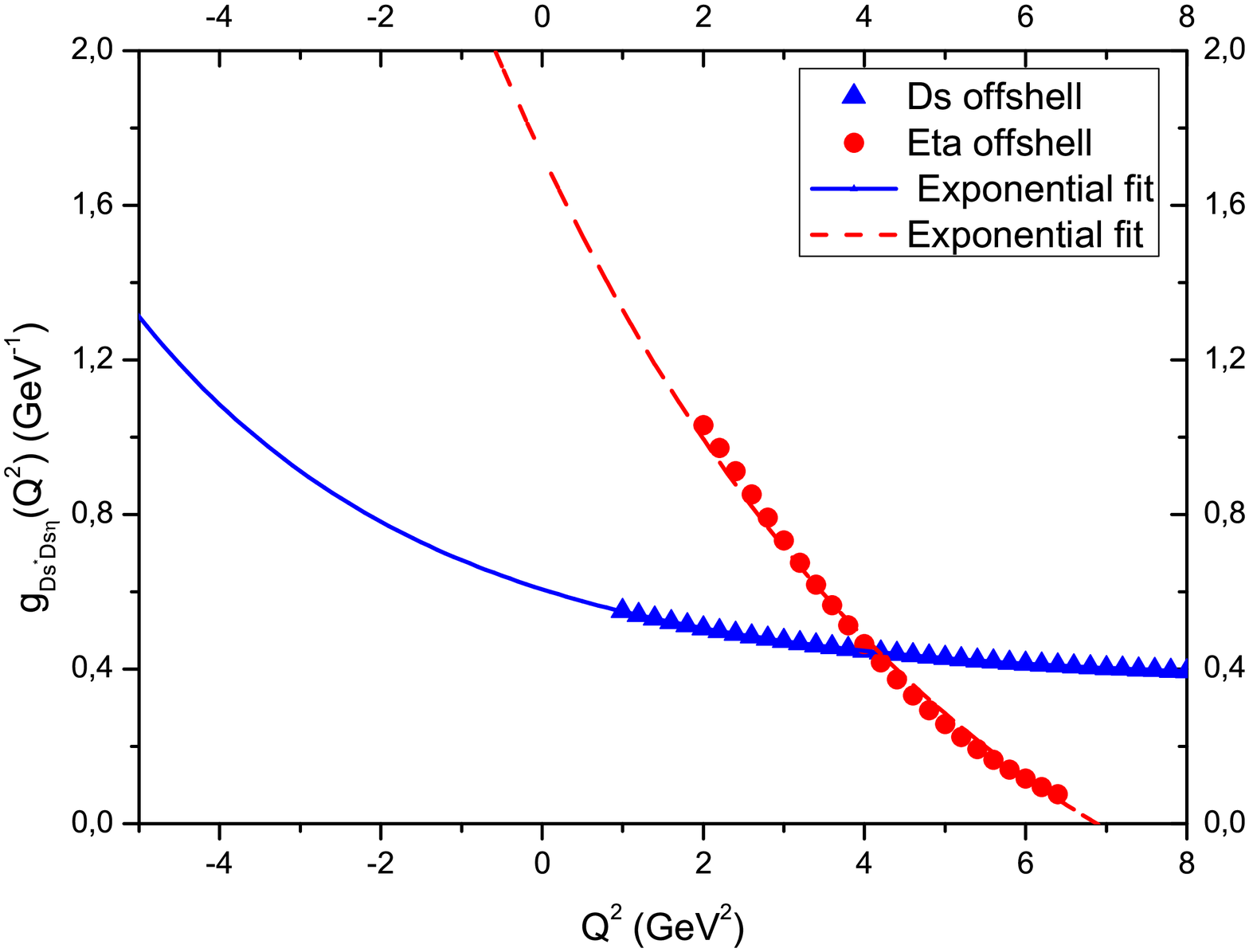}
\includegraphics[totalheight=7cm,width=8cm]{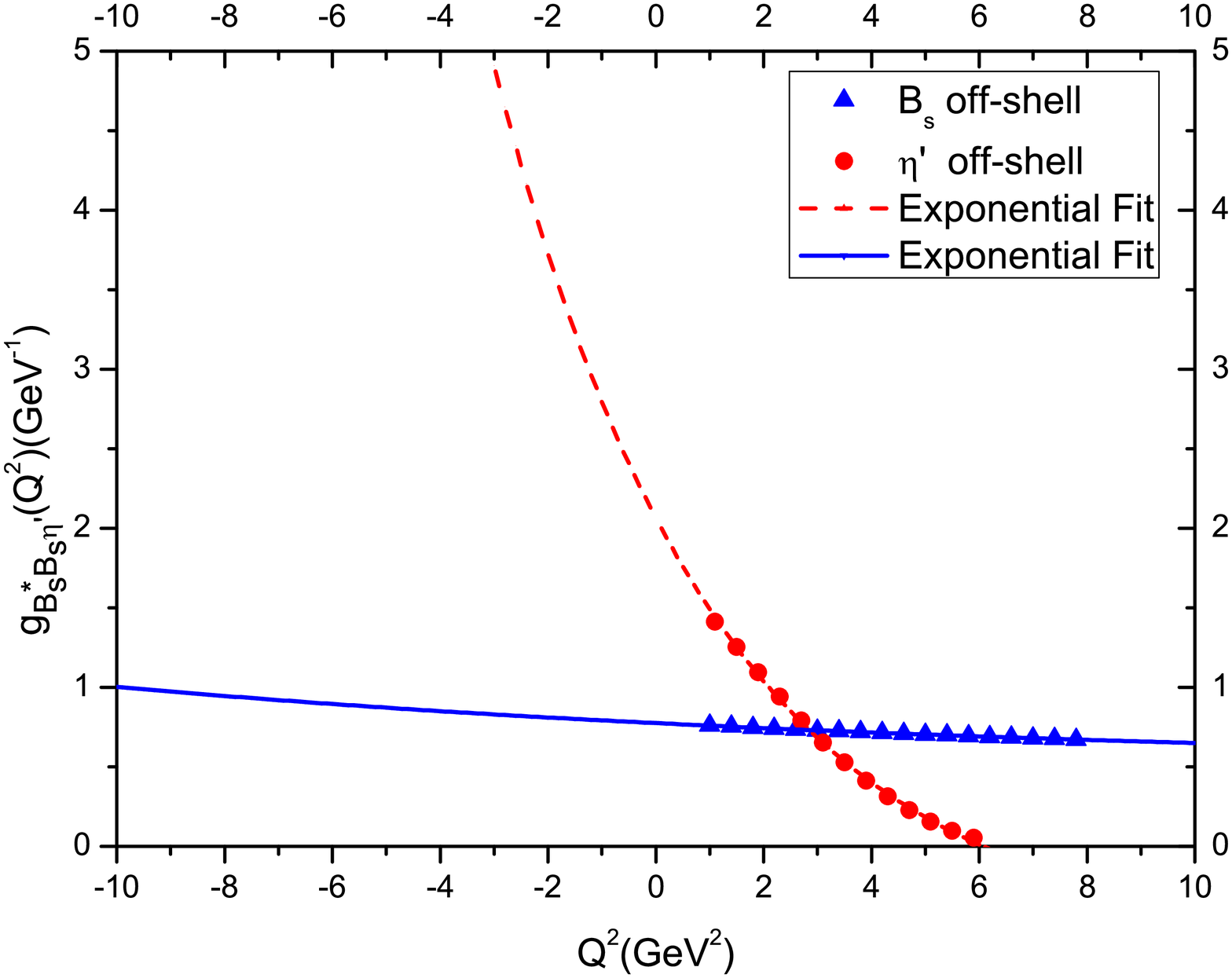}
\caption{\textbf{Left:} $g_{D_s^*D_s\eta}(Q^2)$ as a
function of  $Q^2$. \textbf{Right:} $g_{B_s^*B_s\eta^{\prime}}(Q^2)$ as
a function of $Q^2$. } \label{gQsq}
\end{figure}
\begin{table}[h]
\renewcommand{\arraystretch}{1.5}
\addtolength{\arraycolsep}{3pt}
$$
\begin{array}{|c|c|c|c|}
\hline \hline
 \mbox{ } &
 Q^2=-m_{D_s}^2 &  Q^2=-m_{\eta}^2&  \mbox{Average}

 \\
\hline
  \mbox{$g_{D_s^*D_s\eta}$}
      & 1.06\pm0.24 & 1.86\pm0.36 & 1.46\pm0.30
      \\
    \hline \hline
 \mbox{ } &
 Q^2=-m_{D_s}^2 &  Q^2=-m_{\eta^{\prime}}^2&  \mbox{Average}

 \\
\hline
  \mbox{$g_{D_s^*D_s\eta^{\prime}}$}
      & 0.62\pm0.14 & 0.85\pm0.18 & 0.74\pm0.16
      \\
    \hline \hline
    \hline
 \mbox{ } &
 Q^2=-m_{B_s}^2 &  Q^2=-m_{\eta}^2&  \mbox{Average}

 \\
\hline
  \mbox{$g_{B_s^*B_s\eta}$}
      & 4.54\pm0.90& 6.04\pm1.22 & 5.29\pm1.06
      \\
    \hline \hline
 \mbox{ } &
 Q^2=-m_{B_s}^2 &  Q^2=-m_{\eta^{\prime}}^2&  \mbox{Average}

 \\
\hline
  \mbox{$g_{B_s^*B_s\eta^{\prime}}$}
      & 2.06\pm0.42 & 2.73\pm0.54 & 2.29\pm0.48
      \\
    \hline \hline
\end{array}
$$
\caption{The values of the  coupling constants in $GeV^{-1}$ unit.}
\label{CouplingConstNumericValue}
\renewcommand{\arraystretch}{1}
\addtolength{\arraycolsep}{-1.0pt}
\end{table}

    \end{document}